\documentclass[12pt]{iopart}

\usepackage{graphicx}
\usepackage[breaklinks=true,colorlinks=true,linkcolor=blue,urlcolor=blue,citecolor=blue]{hyperref}

\begin{document}

\title{Influence of clouds on the parameters of images measured by IACT at very high energies}

\author{Dorota Sobczy\'{n}ska}
\address{University of {\L }\'{o}d\'{z}, 
Department of Astrophysics, Pomorska 149/153, 90-236 {\L }\'{o}d\'{z}, Poland}
\ead{dsobczynska@uni.lodz.pl}

\author{W{\l}odek Bednarek}
\address{University of {\L }\'{o}d\'{z}, 
Department of Astrophysics, Pomorska 149/153, 90-236 {\L }\'{o}d\'{z}, Poland}
\ead{bednar@uni.lodz.pl}

\begin{abstract}

Observations with the Cherenkov telescopes are in principle limited to the clear sky conditions due to significant absorption of Cherenkov light by clouds. If the cloud level is high enough or the atmospheric transmission of the cloud is high, then high energy showers (with TeV energies) can still produce enough Cherenkov photons allowing detection by telescopes with large sizes and cameras with large field of view (FOV). In this paper we study the possibility of observations of showers, induced by high energy particles in the atmosphere, in the presence of clouds which are completely or partially opaque for Cherenkov radiation. 
We show how the image parameters of the Cherenkov light distribution on the telescope camera are influenced for different opacity and altitude of the cloud. By applying the Monte Carlo simulations, we calculate the scaled LENGTH and WIDTH parameters with the purpose to separate $\gamma$-ray and proton initiated showers in real data. We show, that the high level of the night sky background effects
the selection efficiency of the $\gamma$-ray initiated showers. However, application of the higher image cleaning level significantly improves expected quality factors. The estimated $\gamma$-ray selection efficiency for the detector with the camera FOV limited to 8$^{\circ}$ is slightly better than for the camera with unlimited FOV, although the number of identified $\gamma$-ray events is lower. We conclude that large Cherenkov telescopes with large FOV cameras can be used for observations of very high energy $\gamma$-rays in the presence of clouds. Consequently, the amount of useful data can be significantly enlarged.

\end{abstract}


\noindent{\it Keywords}: $\gamma$-rays: general -- Methods: observational -- Instrumentation: detectors -- Telescopes 


\submitto{\JPG}

\maketitle

\section{Introduction}

The ground-based $\gamma$-ray astronomy has developed very fast since 1989,
when the first TeV $\gamma$-ray source (the Crab Nebula) was discovered by the Whipple collaboration \cite{whipple}, using the Imaging Air Cherenkov Telescope (IACT). In order to improve the sensitivity of IACTs and to lower the energy threshold, a few systems of Cherenkov telescopes with large mirror area have been built.
This change allowed to view stereoscopically the atmospheric showers improving significantly the sensitivity and the energy threshold. As a result, more than a hundred of new TeV $\gamma$-ray sources at energies above $\sim$100 GeV have been discovered.
The measurement technique is based on the detection of Cherenkov light produced in the atmosphere by the charged relativistic particles from an Extensive Air Shower (EAS). The Cherenkov photons are reflected and focused by the telescope's mirror and finally recorded by the camera (a matrix of photomultipliers mounted in the focal plane of the telescope). The two dimensional angular distribution of the Cherenkov light appears on the camera as the shower image. The image parameters depend on the type and energy of the primary particle, on the distance between the shower axis and the telescope axis (the so called impact parameter) and on the atmospheric conditions.
The number of the registered hadron induced events (the so called background) is several orders of magnitude larger than the number of the registered $\gamma$-ray events from the source. A method to select $\gamma$-rays out of the background created by hadron initiated showers was proposed in 1985 by Hillas \cite{hillas}. This method determines the set of parameters describing the shape and the orientation of the image.

The atmosphere is an integral part of the detector thus should be monitored during data taking. Recent experiments with very large reflectors, such as HESS \cite{aha06}, MAGIC \cite{ale12}, VERITAS \cite{weekes2002,hold11}, monitor the atmospheric conditions in order to improve the data quality \cite{hahn14, gaug13}. The Cherenkov Telescope Array (CTA) Collaboration {\cite {actis11}} also plan to build additional instruments to measure the transparency of the atmosphere \cite{doro13}.

The presence of the clouds during the observations affects the data because the Cherenkov photons are absorbed and scattered by cloud molecules. This causes two important effects in the data.
At first, the detection rate decreases due to reduced amount of Cherenkov light which reaches the telescope. The amount of light can become too small to trigger the telescope. Therefore the effective collection area decreases and the energy threshold increases \cite{rult13}. At second, the images of showers are deformed, since they contain reduced amount of the Cherenkov light. As a result, the reconstruction of the shower can become worse leading to less effective $\gamma$/hadron separation and a degradation of the energy resolution.
The strength of both effects depends on the altitude and transparency of the cloud and on the primary energy of particle. The density of Cherenkov light on the ground is correlated with the height of creation of Cherenkov photons in the $\gamma$-ray initiated shower \cite{sob2009}. Therefore, the effect of clouds on shower images strongly depend on the impact parameter of the shower.

The impact of the atmospheric parameters on the IAC technique has been already studied in \cite{bern00, bern13}. The different atmospheric profiles and atmospheric extinctions, obtained using MODTRAN code, has been applied to the simulations of the Cherenkov light. It has been shown, that the lateral density distribution of the Cherenkov light strongly depend on the atmospheric profile. The expected total signals from aerosol scattered light and Rayleigh scattered light are approximately 2 and 3 orders of magnitudes smaller than from the direct Cherenkov light for proton showers with primary energy of 100 TeV. The influence of additional absorption due to the opaque cloud presence was not investigated in \cite{bern00, bern13}.

A correction method of the reconstructed energy of the shower and fluxes of primary particles due to the presence of low-level aerosols have been applied to the HESS data \cite{nolan10}. Two important effect have been discussed in this paper. First, in the presence of low-level aerosols the reconstructed energy is biased towards lower values if the Monte Carlo simulations of a clear sky are used. Second, the effective correction area (as a function of reconstructed energy) is reduced for the simulations with additional absorption by aerosols. As a result, the simulations of the atmospheric parameters, measured by LIDAR system \cite{bour13}, have to be applied for the reconstruction energy and flux determination. The correction method described in \cite{nolan10} has been applied to the primary energies below 10 TeV using MODTRAN code to simulate the atmosphere with low-level aerosols.

It has been shown in \cite{dorn09} that the data taken by MAGIC during nights with additional partial absorption by Saharan Air Layer (below 5.5 km) can be corrected when the absorption is lower then 40$\%$ for energies below 1 TeV. The shower maximum is well above the dust layer for those energies. The additional factor was applied in the absolute light calibration in order to compensate for the additional absorption. The image parameter called SIZE (this is the sum of all signals from pixels which make the image) has been corrected to get properly reconstructed primary energy. Simple correction method of the reconstructed energy and flux has been applied to the MAGIC stereo data \cite{fruck13}. At first, the reconstructed energy is scaled up using the weighted aerosol transmission of the atmosphere. This value is determined from the data taken by the LIDAR. The collection area for the corrected energy corresponds to the energy before scaling up. It has been shown that the flux of the $\gamma$-ray source can be reproduced by using this method.

{The reconstructed energy and the flux can be corrected for the low-layers of aerosols (or dust). They are below the shower maximum of the relatively low energy particles (lower than approximately 10 TeV). The clouds can also influence showers above or close to the shower maximum (depending on the primary energy and the altitude of clouds). This may lead to the significant changes of the shower image. It has been shown \cite {sob13}, that for fixed energy of primary particle, images can be strongly affected by fully opaque clouds.

In this paper we investigate in detail the influence of cloud parameters (its altitude and transmission) on Hillas parameters which describe the shape and the position of image of the shower. The aim of our paper is to show the ability of the $\gamma$-rays selection using the Imaging Air Cherenkov Technique for the very high energies.
Based on the Monte Carlo (MC) simulations, we show how the presence of clouds change the lateral distribution of the Cherenkov photon densities for $\gamma$-ray and proton vertical showers. We show the distributions of the image parameters (the scaled WIDTH and LENGTH \cite{daum97}) for clouds with various transmissions (T) and altitudes in cases of the primary proton and $\gamma$-ray showers at very high energies.
These parameters are applied for the separation of the $\gamma$-ray and hadron induced showers. The quality factor has been used to demonstrate the $\gamma$-ray selection efficiency from the hadronic background. We also study the influence of different elements on the data analysis procedure (i.e. the scaling procedure, the level of image cleaning and the level of the night sky background) on the quality factors. 
Finally, we consider the possibility of the $\gamma$/hadron separation at very high energies in the presence of clouds for cameras with FOV limited to 8$^{\circ}$. Cameras with such FOV are proposed to be applied in the middle size telescopes (MST) in the CTA project. We conclude that large mirror Cherenkov telescopes with large FOV camera should allow detection of very high energy $\gamma$-ray showers even with the presence of relatively low altitude clouds.

\section{Description of the Monte Carlo simulations}

In order to calculate the Cherenkov light distribution on the ground we simulate the development of vertical showers initiated by $\gamma$-rays and hadrons in the Earth's atmosphere by using the CORSIKA code (version 6.99 \cite{heck,knapp}). The UrQMD {\cite{urq1,urq2}} and QGSJET-II {\cite{ostap06a,ostap06b}} interaction models have been applied for the the low and high energy ranges. The US standard atmosphere profile has been used.  Rayleigh and Mie scattering of the light in the atmosphere were also taken into account according to the Sokolsky formula \cite{sokol}. 
Simulations have been performed for two sets of vertical showers. In the first set, we simulate the $\gamma$-ray and proton induced showers at fixed energies of 5, 10 TeV and 10, 20 TeV, respectively. For those energies, we calculate the average longitudinal and lateral distributions of the Cherenkov light at the MAGIC site \cite{ale12} (the altitude of 2.2 km a.s.l.), with and without the presence of fully opaque clouds. 
Densities of Cherenkov photons for the cloud at fixed altitudes, between 3 km to 10 km a.s.l., were obtained. To determine the Cherenkov light densities, we apply the telescope with the mirror area of 225 m$^{2}$, with unlimited camera FOV and FOV limited to 8$^{\circ}$. The results of MC simulations base on the 200 simulated showers, except for the case of showers initiated by protons with energies 20 TeV in which case we consider only 100 simulated showers.
The second set of simulations concerns the angular distribution of Cherenkov photons in cases of eight different impact parameters (between 32.5 m and 312.5 m) and for the same detector area. 
The histograms of angular distributions were obtained by counting the Cherenkov photon in the ranges -1$^o$ and 14$^o$ (in the x axis) and -7.5$^o$ to 7.5$^o$ (in the y axis). The width bin of the histogram was set to 0.1$^o$, what is approximately the typical pixel size in operating IACT. 

The simulations have been performed without the presence of any clouds and with clouds at the different altitudes equal to 10, 7, 6 and 5 km a.s.l.. We performed simulations assuming different transmission of the cloud, starting from T=0 (opaque) up to T=1 (transparent) with the step equal to 0.2. 
Cherenkov photons produced in the cascades above the cloud level were partially absorbed with the probability equal to the cloud transmission.
Moreover, we added to the Cherenkov light photons from the night sky background (NSB) with the level of $1.75 *10^{12} ph/(m^2~sr~s)$ (corresponding to the background measured on La Palma \cite{nsb}). 
We have simulated showers initiated by vertical protons with energies between 4.32 TeV and 100 TeV and the differential spectral index equal to -2.73. The vertical $\gamma$-ray showers have been simulated in the energy range between 2 TeV and 100 TeV. The differential $\gamma$-ray spectral index was chosen to be -2.6, which is close to the index of the Crab Nebula spectrum for energies above 300 GeV \cite{wagner} and 500 GeV \cite{aha2004}. We have simulated altogether 13000 $\gamma$-ray and proton showers for each atmospheric condition described by above defined parameters of clouds.
It is worth to mention here that we decided not to add any features of the telescopes to the simulations because we wanted to know how the image parameters change only with the presence of clouds (without any additional limitations due to e.g. the Quantum Efficiency of photomultipliers or the trigger conditions). The only used limitations are related to the localization and the size of the reflector and to the camera FOV. 

\section{Results and discussion}
\subsection{Longitudinal distribution and lateral density distribution of $\gamma$-ray showers}

At first, we analyse the results of simulations of $\gamma$-ray and proton induced showers for their fixed primary energies. 
We show the average number of Cherenkov photons, produced in the atmosphere within layers with thickness of 10 $g/cm^2$, versus the atmospheric depth for energies 5, 10 TeV of primary $\gamma$-rays and 10, 20 TeV of primary protons (see Fig.~1). All longitudinal distributions for proton induced showers are multiplied by a factor of 10 for better visibility. In the same Figure, we also plot the longitudinal distribution of Cherenkov photons as seen by the camera with FOV limited to 8$^{\circ}$. Four vertical lines (thin solid) correspond to the clouds at altitudes 10, 7, 6 and 5 km a.s.l.. We show that most of the light is produced below 10 km for all simulated energies. 
Note that, in the case of limited FOV of camera, the shower maximum is higher in the atmosphere for both types of simulated primary particles. Below the shower maximum, the amount of Cherenkov light in the camera with 8$^{\circ}$ FOV is a factor of $\sim$2 lower in comparison to all Cherenkov light produced in the cascades. This results in a significant reduction of the light density on the ground. The shower maximums, for the investigated energy range of primary particles, lay between 5 to 10 km above the sea level. Therefore, the presence of clouds below this height should strongly effect the Cherenkov light arriving to the telescope.

\begin{figure}[t]
\begin{center}
\includegraphics*[width=12cm]{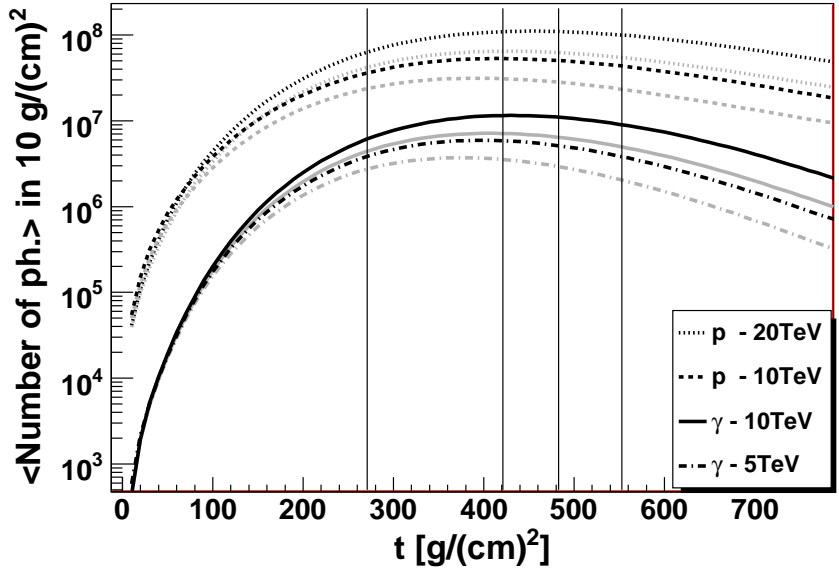}
\end{center}
\caption{The mean number of Cherenkov photons produced at different atmospheric depth for proton showers with energies 20 TeV (dotted lines) and 10 TeV (dashed lines) and $\gamma$-ray showers with energies 10 TeV (solid lines) and 5 TeV (dashed-dotted lines). Black lines show the number of all Cherenkov photons created by secondary particles of the shower, while grey lines show the number of Cherenkov photons within the 8$^{\circ}$ FOV camera. The longitudinal distributions for proton induced showers are multiplied by a factor of ten to allow easier comparison. Four vertical lines correspond to the altitudes of 10, 7, 6 and 5 km a.s.l.}
\label{longi}
\end{figure}

In general, densities of Cherenkov light on the ground and heights of Cherenkov photon production are correlated \cite{sob2009}. Therefore, the decrease of the Cherenkov photon density, due to the presence of cloud, should depend on the impact parameter of the shower for the fixed cloud transmission and altitude. In order to check this effect quantitatively, we show how the fully opaque clouds can reduce the average Cherenkov light density for different energies of primary particles, 5 TeV $\gamma$-ray showers (a) and 20 TeV proton showers (b) (see Fig.~2). Two cases are considered, i.e. the case of camera with unlimited FOV (black lines) and limited FOV to 8$^{\circ}$ (grey lines). The figures show the ratio between the absorbed and not absorbed Cherenkov photon densities (the density fraction) as a function of the impact parameter of the shower. As expected, this ratio increases with the altitude of the fully opaque cloud. 

The impact parameter of the shower is defined as the distance between the shower axis and the telescope axis. 
Most of the Cherenkov photons produced very high in the atmosphere hit the ground close to the shower axis. Therefore, the presence of clouds significantly reduce expected Cherenkov photon densities at very small impact parameters. Note, that all Cherenkov photon density distributions (Fig.~2) have the local minimum at the impact parameter R=0 m, except the case of $\gamma$-ray with the cloud at 3 km a.s.l.. The second local minimum in these curves (at the impact parameter of $\sim$110 m, i.e. so called the hump position) are related to the shower maximum. For larger impact parameters, the fraction of unabsorbed photons slowly decreases with the impact parameter. For the cloud at the altitude of 10 km, the loss of the light due to the total absorption in the cloud is only $\sim$10$\%$. Therefore, we expect relatively small changes in the image parameters of showers for the case of high altitude clouds.
The limited camera FOV strongly influences the density fraction of proton and $\gamma$-ray initiated showers for the impact parameters larger than the hump position.
 The density fraction of the Cherenkov light within 8$^{\circ}$ camera FOV falls with the impact parameter significantly faster than in the case of the detector with unlimited FOV.

\begin{figure*}[t]
\vskip 6.truecm
\includegraphics{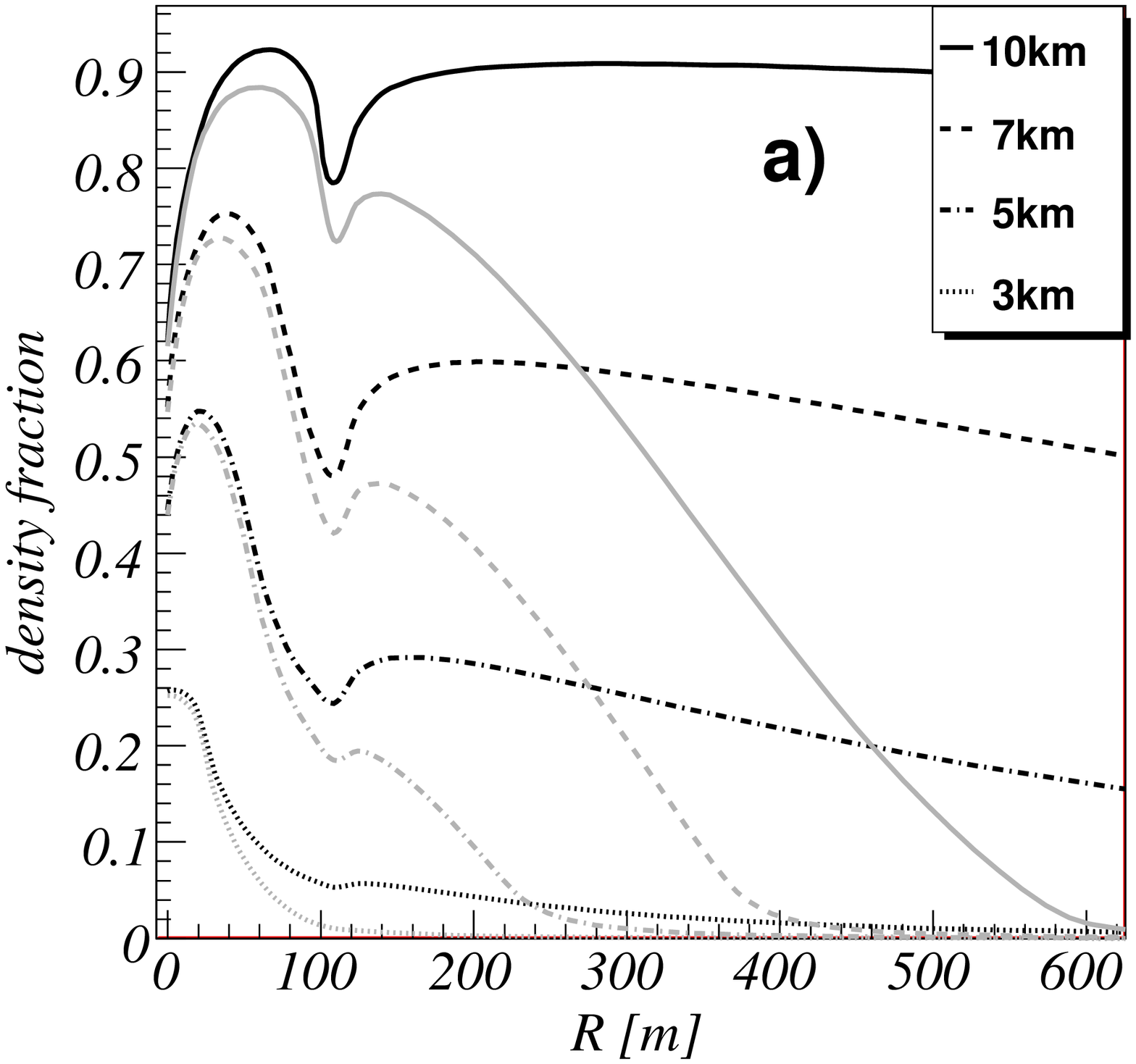}
\includegraphics{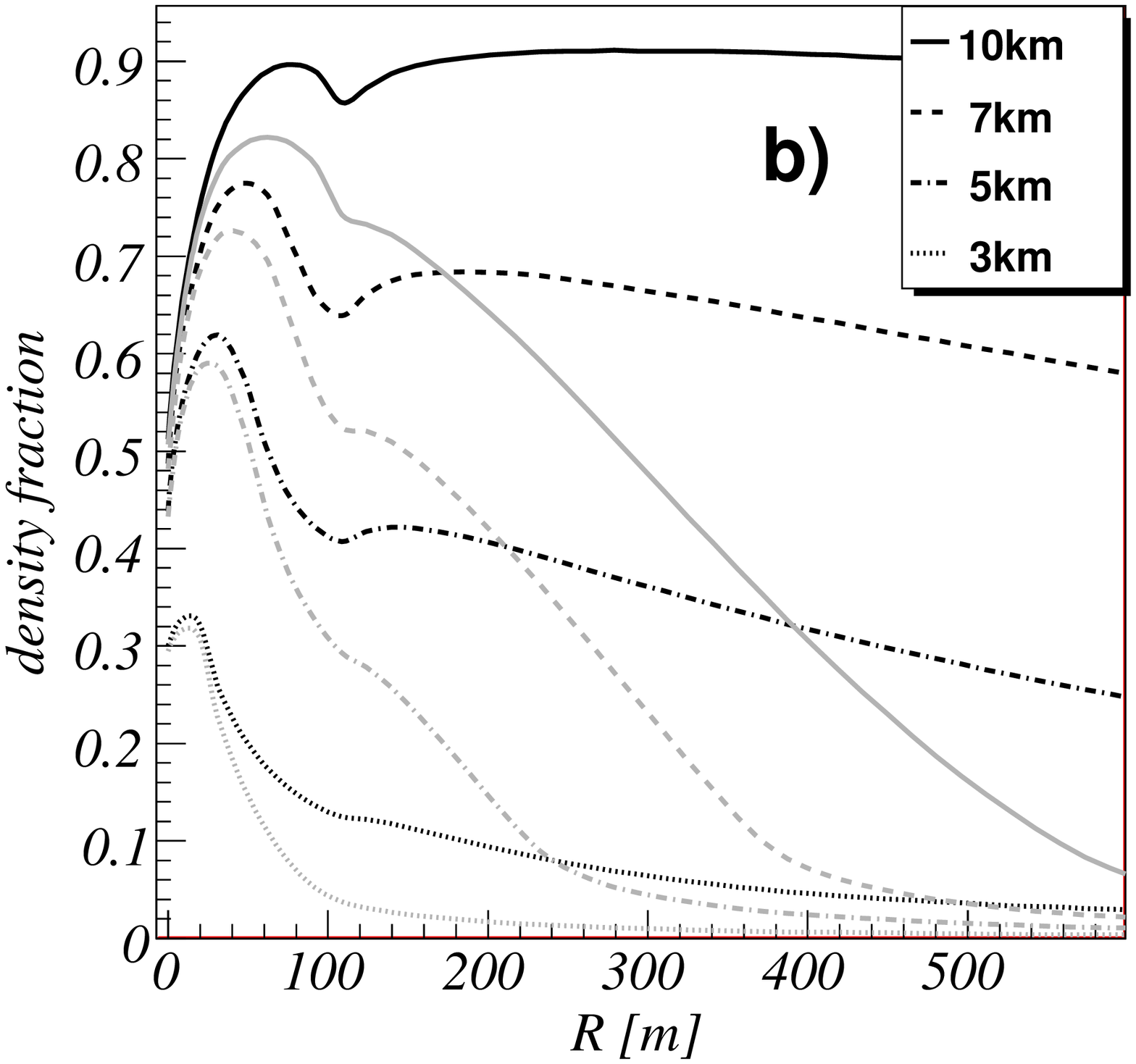}
\caption{The ratio between the Cherenkov photon densities, which are absorbed by the cloud and without taking into account such absorption, versus the impact parameter of the primary particle assuming fully opaque clouds at 10 km a.s.l. (solid lines), 7 km (dashed lines), 6 km (dashed-dotted lines) and 5km (dotted lines). The results for the detector with unlimited FOV (black curves) and for FOV of camera limited to 8$^{\circ}$ (grey curves). The $\gamma$-ray initiated showers with energy 5 TeV (a) and proton initiated showers with energy 20 TeV (b) are considered.}
\label{dens_fract}
\end{figure*}

\subsection{Shower images and their parameters}

We have performed simulations of $\gamma$-ray and proton initiated showers in order to investigate the images of the Cherenkov light produced by these showers in the presence of clouds. One thousand $\gamma$-ray induced showers, with primary energy equal to 2 TeV, 
and also one thousand proton initiated showers, with energy equal to 10 TeV, have been used in our analysis. 
Simulations were done for the clear sky conditions and for the fully opaque clouds at the altitude of 7, 6 and 5 km a.s.l.. The average angular distributions of the Cherenkov light on the telescope have been obtained for the telescope placed at distance of 152.5 m from the shower core (see Fig.~3). We show that the images with the presence of clouds are shifted outwards of the camera center (defined by x=0 and y=0). For lower cloud altitude, the shift of the image is clearly larger. As expected, these images also contain less Cherenkov photons in comparison to the clear sky simulations (see different ranges of z-axes in Fig.~3). 

Two Hillas parameters called LENGTH and WIDTH (describing the image shape) are defined as the standard deviations of profiles of two dimensional Cherenkov light angular distribution \cite{hillas}. The LENGTH and the WIDTH give the values along the long and short axis of the image, respectively. The DIST parameter (describing the position of the image) is the distance between the image center of gravity and the camera center (or the position of the source on the camera plane in cases of wobble observations) \cite{hillas}. The camera FOV of 8$^{\circ}$ corresponds to the limitation of the x axis to 4$^{\circ}$.

If the additional absorption by cloud is taken into account, then the image parameters change in comparison to images obtained in the clear sky simulations (see Fig.~3). Influence of the limited camera FOV on the image parameters depend on the altitude of the fully opaque cloud. The impact of the limited camera FOV is stronger for the lower altitude of the fully opaque clouds.

\begin{figure}[t]
\begin{center}
\includegraphics*[width=12cm]{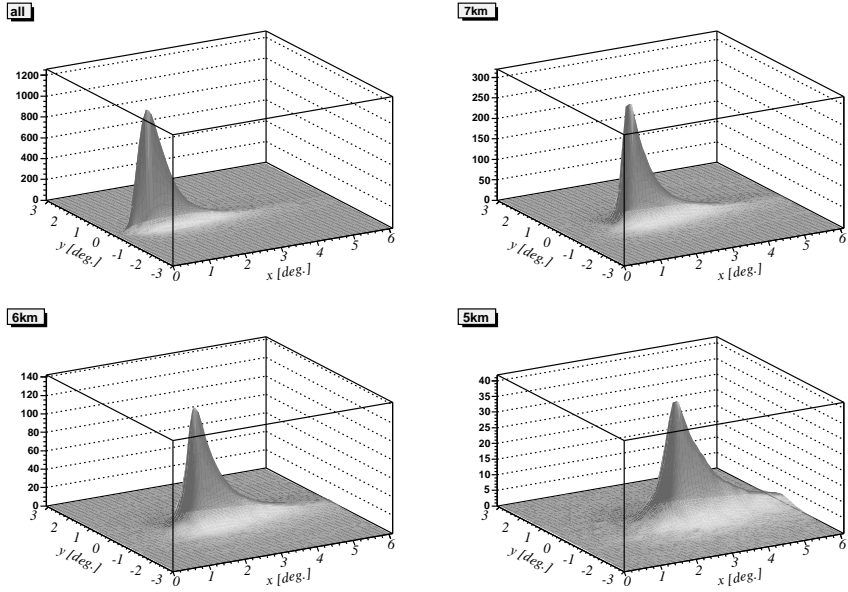}
\end{center}
\caption{Average images of Cherenkov light at the telescope altitude for the impact parameter equal to 152.7 m and for the $\gamma$-ray initiated showers with energy equal to 2 TeV. The simulations with the clear sky conditions are marked by "all". Clouds are fully opaque at the altitudes equal to 7, 6 and 5 km a.s.l.. Note, different values on z-axes.} 
\label{images}
\end{figure}

To obtain the distributions of the DIST, LENGTH and WIDTH parameters, we simulated the primary particles ($\gamma$-rays, protons) with the spectra as described above. The angular distributions of the Cherenkov light, obtained from the MC simulations, have been cleaned. Signals recorded by pixels of the camera include photoelectrons created by both Cherenkov light and NSB photons. Photoelectrons produced by light of the NSB are randomly distributed over all pixels of the camera. Therefore, those photoelectrons should be subtracted before calculating the parameters of shower image. The cleaning procedure of the image is used for this purpose. The results of our simulations also contain photons of the NSB, which are distributed over the whole two dimensional angular distribution of Cherenkov light. In this paper, the image parameters are calculated after applying the cleaning image procedure which neglects pixels with the signal smaller than the chosen level. In our study, we checked three levels of the cleaning: 30, 45 and 60 photons.
Distributions of the Hillas parameters are interpolated in the range of the shower impact parameter between 22.5 m and 322.5 m. 
Note that very high energy $\gamma$-ray initiated showers, close to the telescope axis, create the images, which are usually very difficult to separate from hadron initiated showers since their images form very broad ellipses. Therefore, we limit the impact parameter of the analysed showers to values greater than 22.5 m. On the other hand, for the very large impact parameters, the Cherenkov photon density can be too low to trigger the telescope. Therefore, the upper limit on the impact parameter has been fixed to 322.5 m, 
which is about 200 m farther from the shower axis than the hump position (see Fig.~2).

We show dependencies of the DIST, LENGTH and WIDTH \cite{hillas} on the transparency of the cloud at the level 6 km a.s.l. (see Fig.~4). Both, $\gamma$-ray and proton induced showers, are shown for the images with the SIZE above 2000 photons. In our study, photons are not converted to photoelectrons, so the SIZE is given in photons. 
The distributions of DIST parameter of the $\gamma$-ray and proton initiated showers become wider for lower cloud transmission (see Figs.~4a,d). The images of showers are significantly shifted towards larger values in the case of fully opaque clouds. This is due to the fact, that these images do not contain photons created above the cloud altitude, where the Cherenkov angle is small and the angular distribution of charged particles in the shower is narrow. 

The LENGTH distribution for primary $\gamma$-rays, which reflects the angular distribution of the charged particle along the shower core (Fig.~4b), is shifted towards lower values for the fully opaque clouds. The LENGTH distributions, obtained for higher cloud transparencies, show different tendency. The scale of the shift towards larger values increases with decreasing transmission of the cloud for both $\gamma$-ray and proton initiated showers (see Fig.~4b,e). This feature can be explained by the fact, that for the cloud with lower transparency images are less concentrated and therefore, the LENGTH parameter is larger. 

\begin{figure*}[t]
\vskip 9.5truecm
\includegraphics{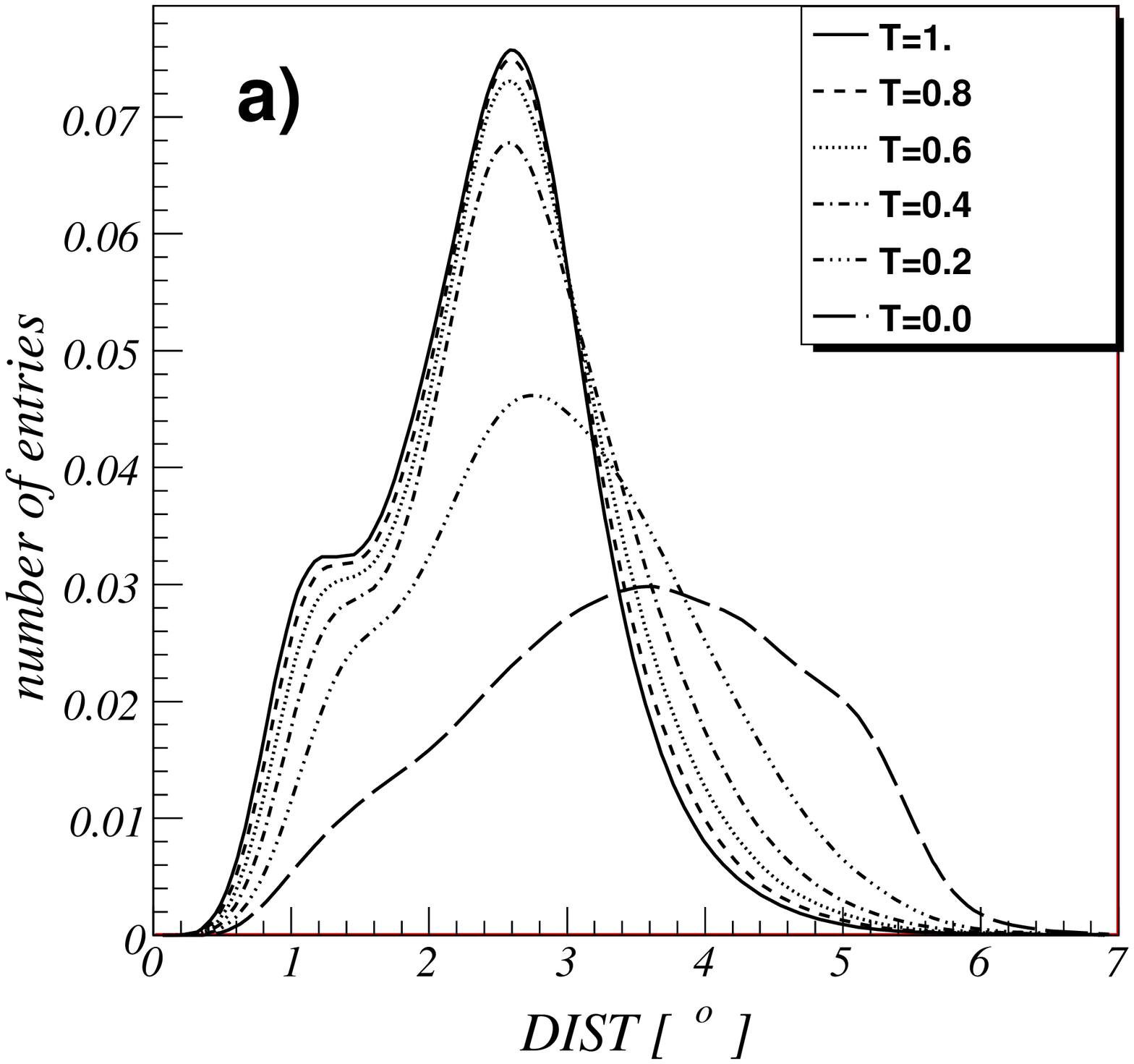}
\includegraphics{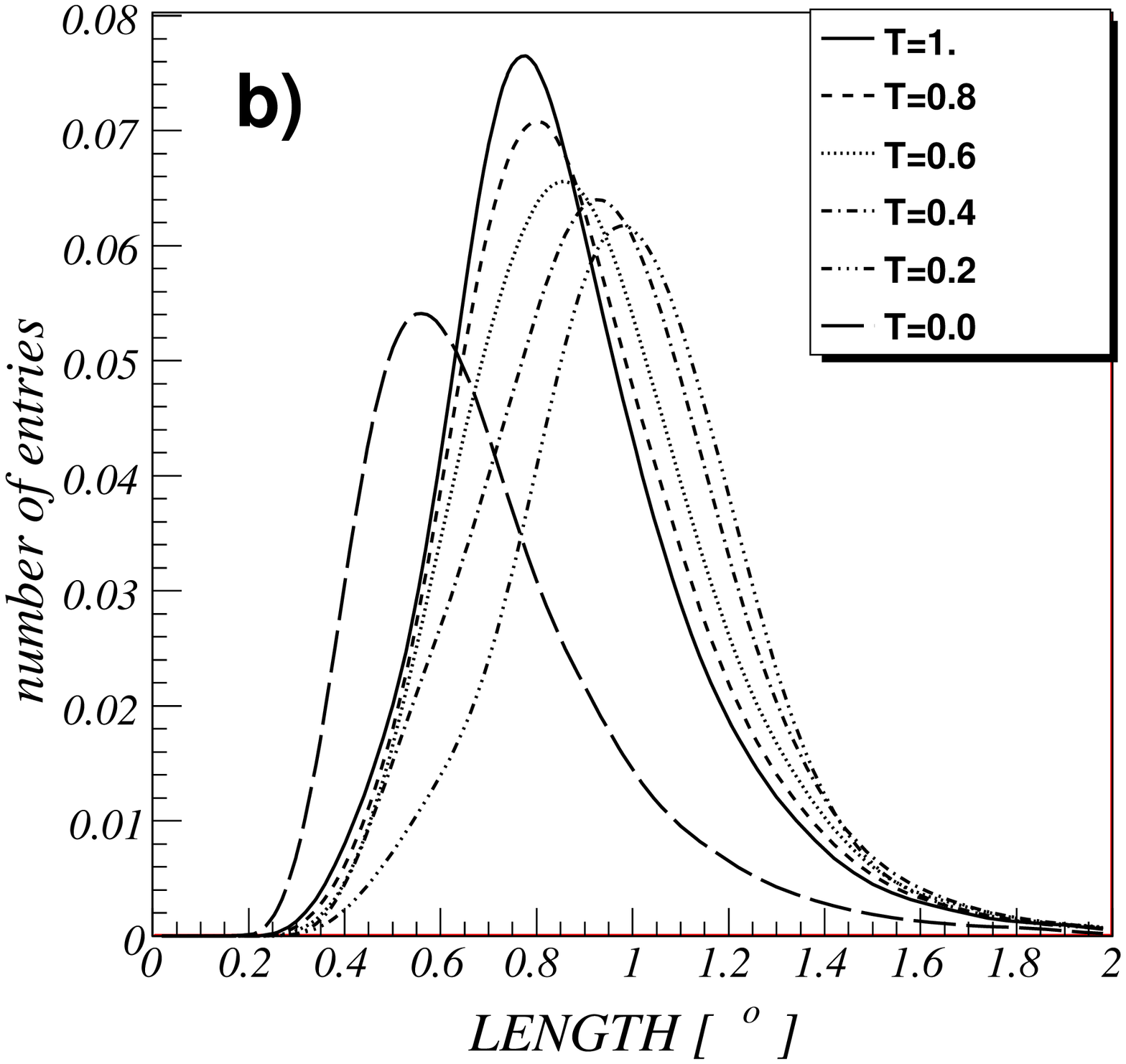}
\includegraphics{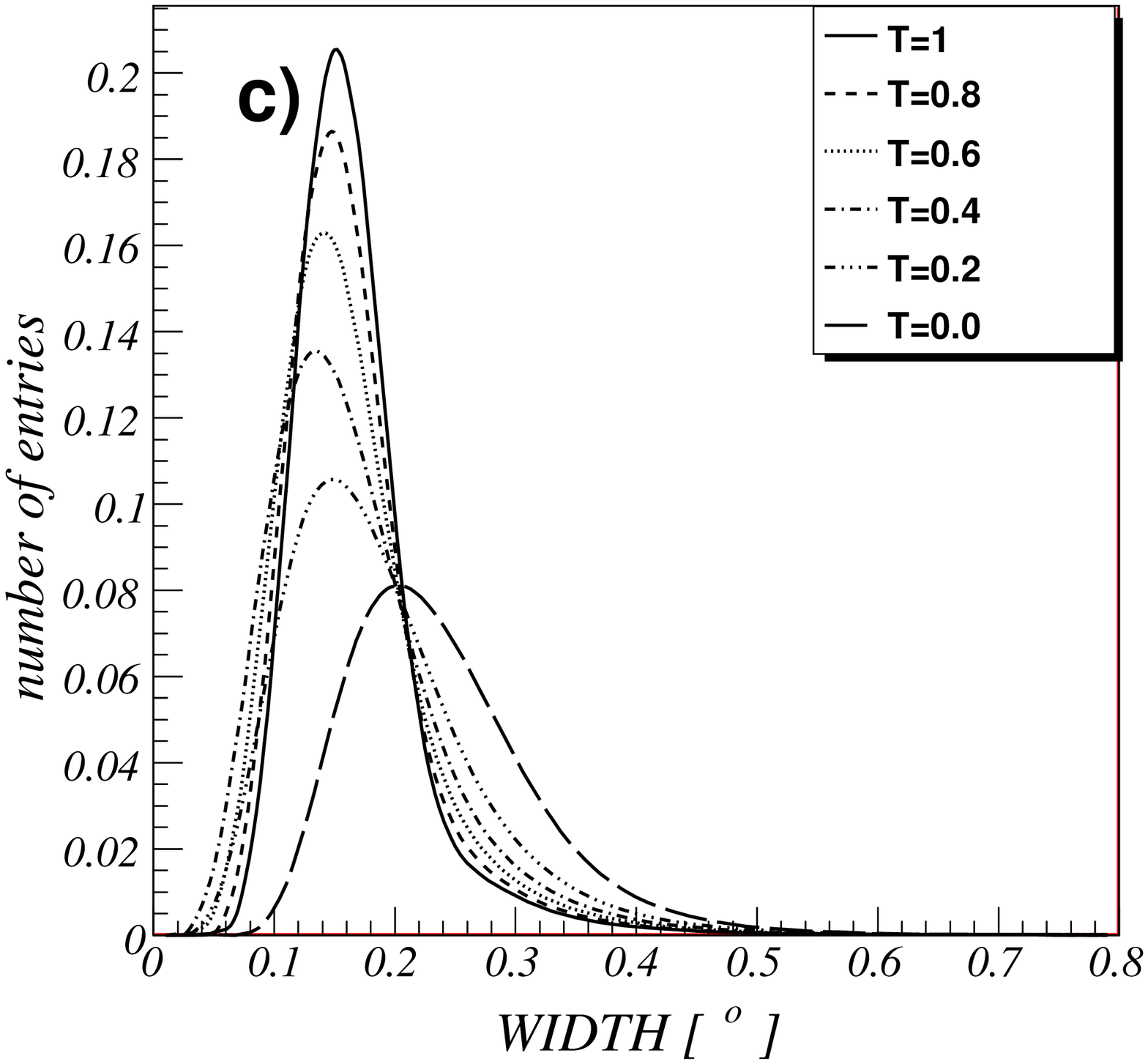}
\includegraphics{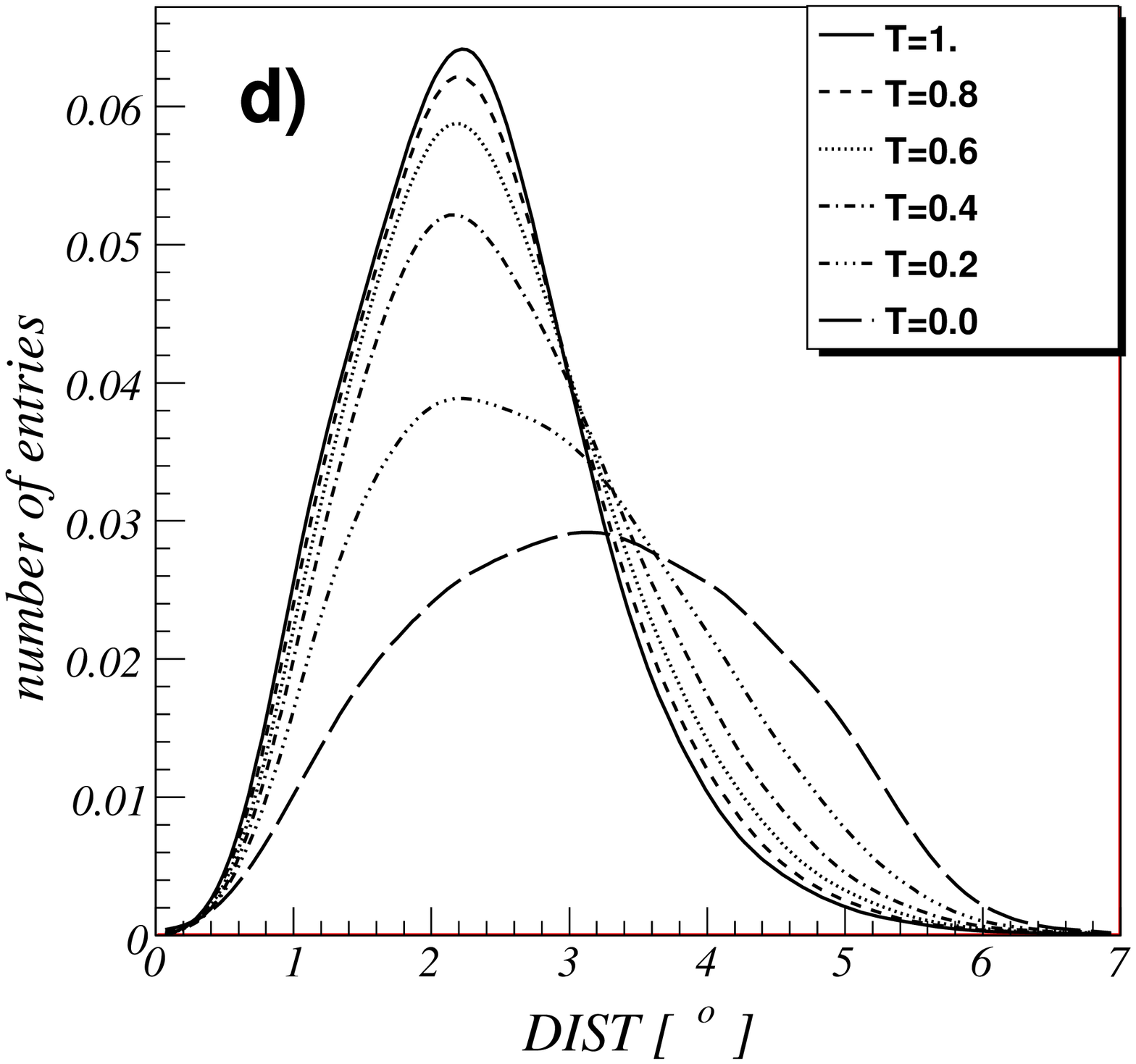}
\includegraphics{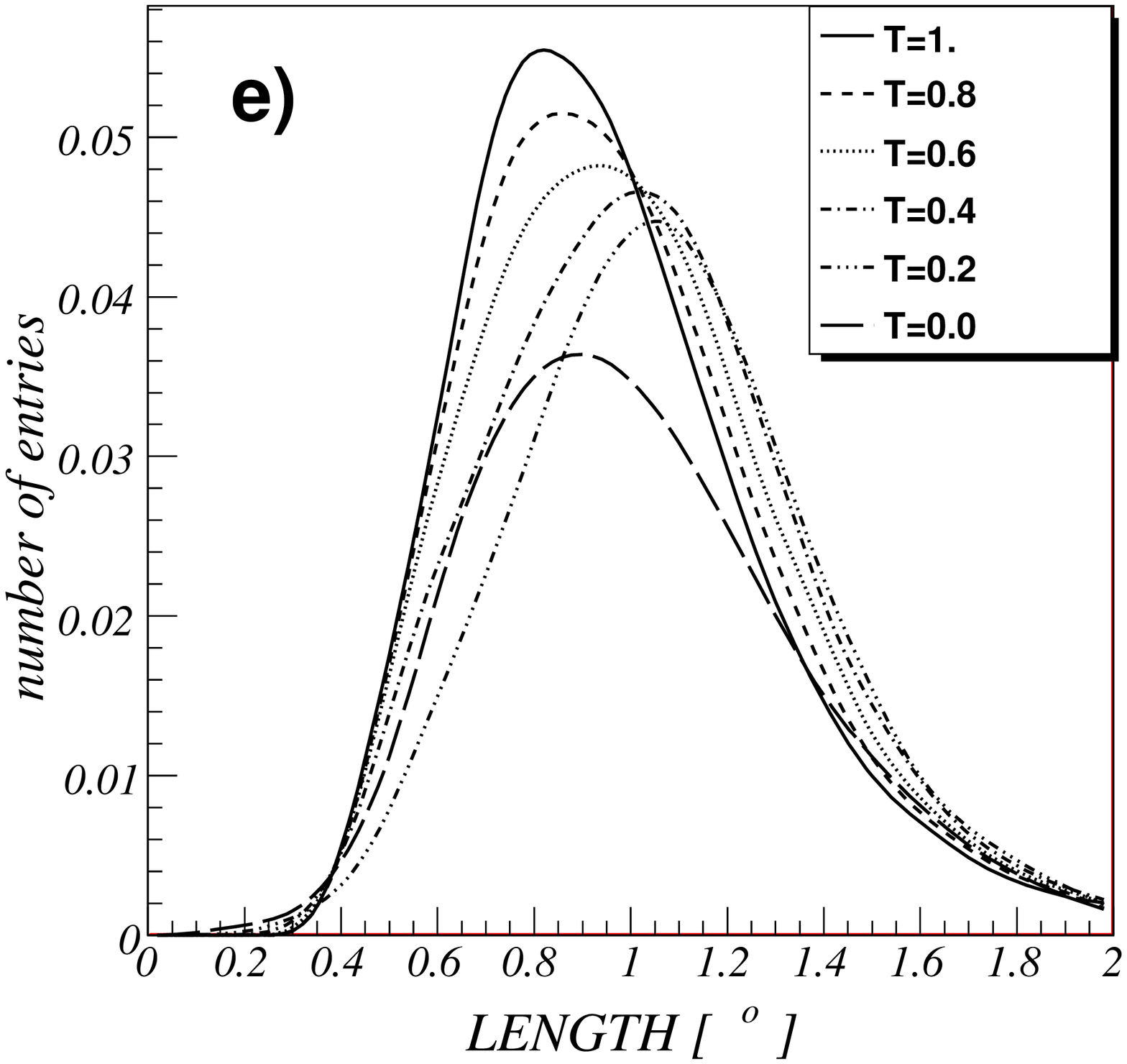}
\includegraphics{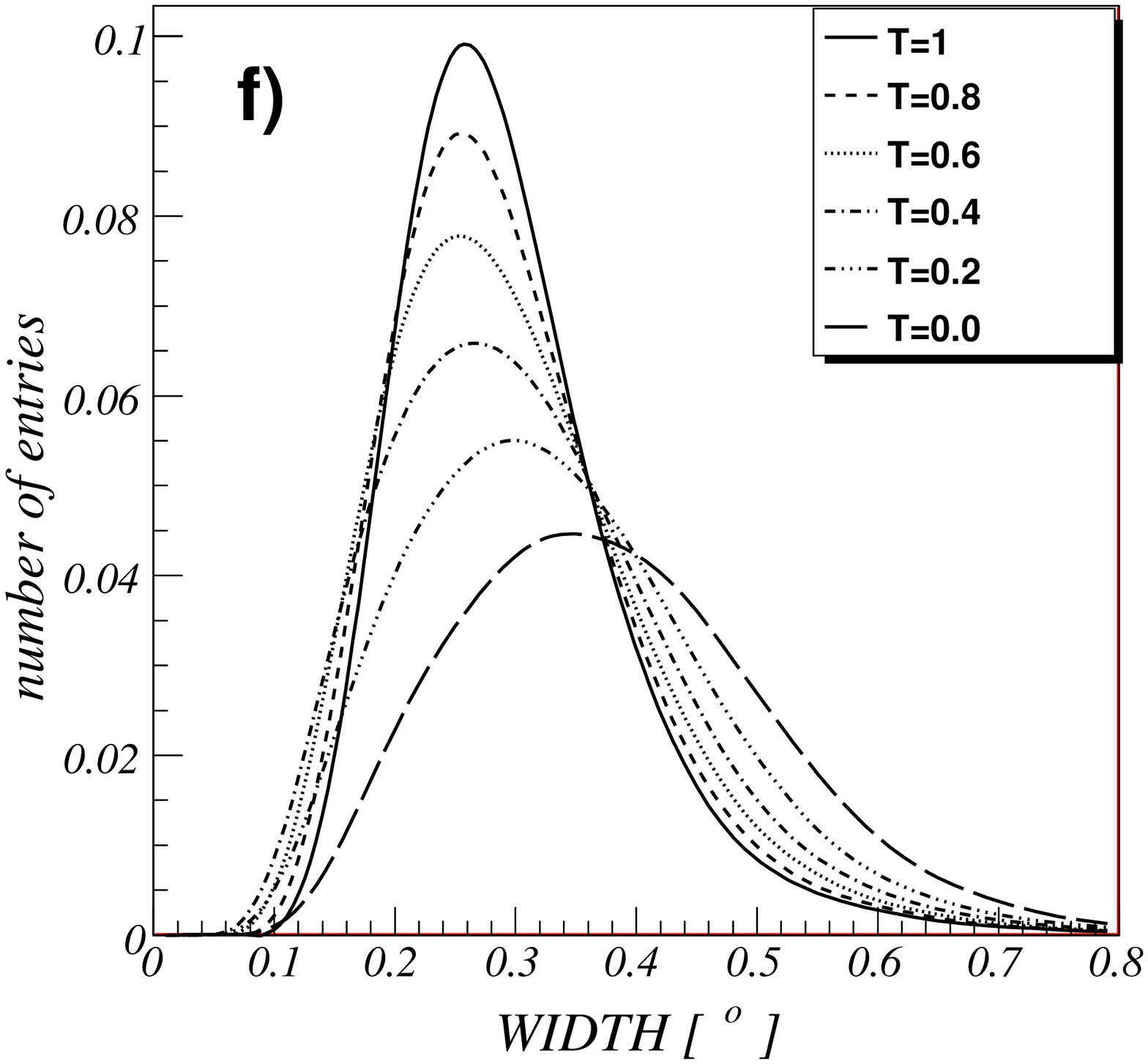}
\caption{The distribution of Hillas parameters for the $\gamma$-ray initiated showers (top panel) and for the proton initiated showers
 (bottom panel) in the case of particles with the power law spectrum as described in the text.
The clouds at the altitude of 6 km a.s.l are characterised by different transmission equal to T = 1 (no absorption of Cherenkov light, solid), 0.8 (dashed), 0.6 (dotted), 0.4 (dashed-dotted), 0.2 (dotted-dotted-dotted-dashed) and 0 (long dashed)}
\label{proton_par}
\end{figure*}

The WIDTH distributions reflect the angular distributions of the charged particles in the direction perpendicular to the shower axis. Our simulations show that in the case of more opaque clouds the WIDTH distributions are wider for both simulated primary particles (see Fig.~4c,f). The WIDTH distributions are shifted towards larger values for the clouds with transmission of 0.2 and 0, in the case of the proton initiated showers and for the transmission equal to T=0 in the case of $\gamma$-ray initiated showers. Those very opaque clouds at relatively low altitude create images much wider than those created under clear sky conditions. 
The changes of the image parameters described above can be noticed in all simulated cloud altitudes. However, in the case of the cloud at 10 km, the differences between clear sky and all investigated transparencies are very small. The presence of clouds influences the 
$\gamma$-ray and proton images in a similar way. Therefore, it is clear that selection of $\gamma$-ray showers is possible at very high energies. 

\subsection{The $\gamma$/hadron separation} 

In this section we analyse the possibility of selection of $\gamma$-ray initiated showers from the background of proton initiated showers in the presence of clouds by using the Hillas parameters describing the image shape only. Those Hillas parameters, which describe the image orientation, can not be used in our study because proton initiated showers were simulated from the direction of the $\gamma$-ray source. The $\gamma$/hadron separation method, based on simply static Al cuts in WIDTH and LENGTH, does not take into account the effect of correlation of those parameters with the image SIZE. Other variables, so called scaled WIDTH (WIDTH$_{S}$) and scaled LENGTH (LENGTH$_{S}$), have been proposed as better indicators of the $\gamma$-ray initiated showers, since they are independent on the image SIZE (for definition see \cite{daum97}). The scaling factors are calculated from simulated $\gamma$-ray images only. They are applied to the calculations of the WIDTH$_{S}$ and LENGTH$_{S}$ for the images of $\gamma$-ray and proton initiated showers. As a result, the WIDTH$_{S}$ and LENGTH$_{S}$ distributions of the $\gamma$-ray initiated showers are gaussian distributions with mean equal to 0 and dispersion equal to 1. Both main statistical parameters describing WIDTH$_{S}$ and LENGTH$_{S}$ distributions for proton images have larger value than for $\gamma$-ray images. The WIDTH and the LENGTH are always positive and they are measured in degrees, while the WIDTH$_{S}$ and the LENGTH$_{S}$ can also be negative and they do not have units. The parameters WIDTH$_{S}$ and LENGTH$_{S}$ allow to optimize the cut limits independent on the SIZE parameter.

\begin{figure*}[t]
\vskip 6.truecm
\includegraphics{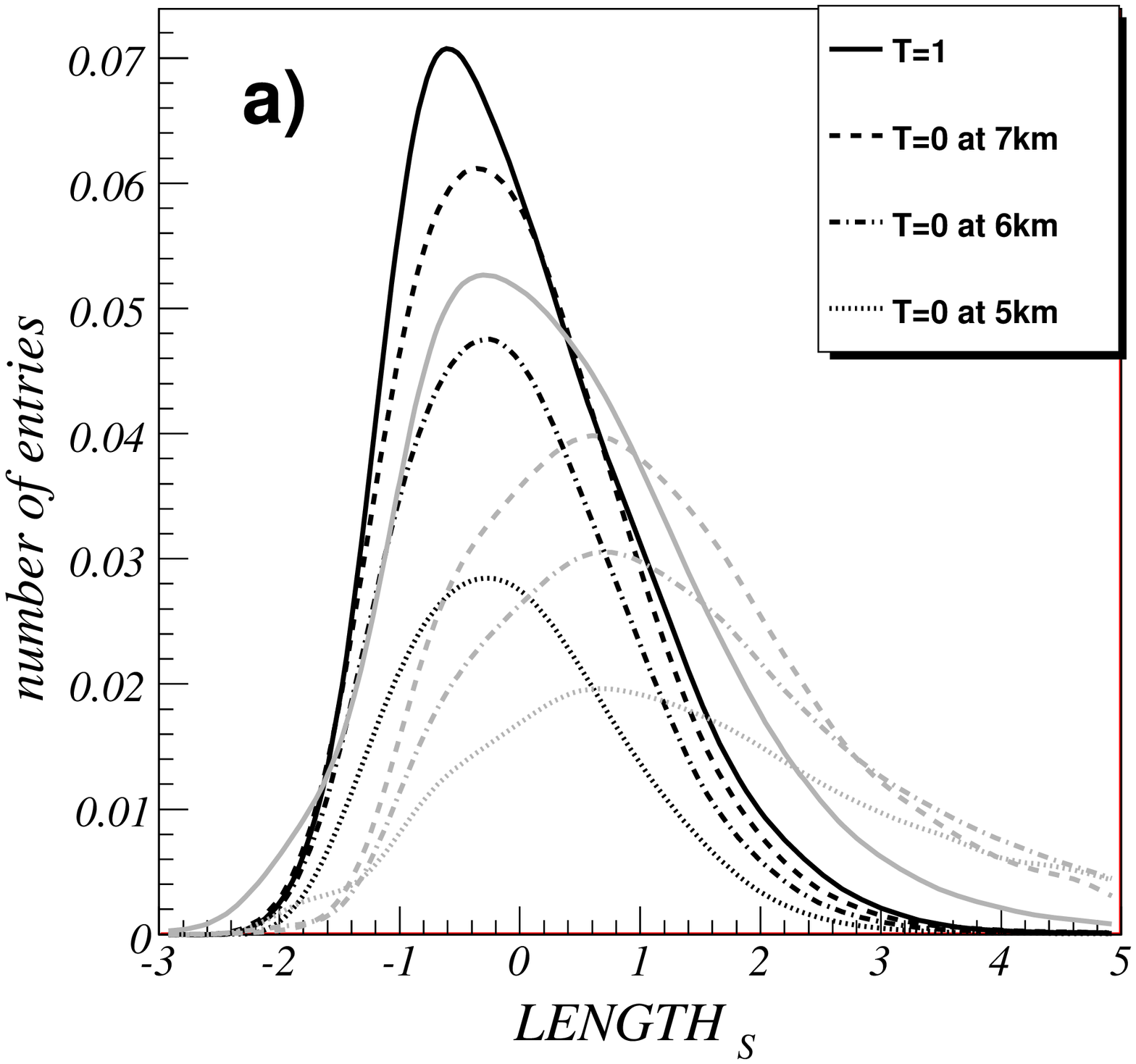}
\includegraphics{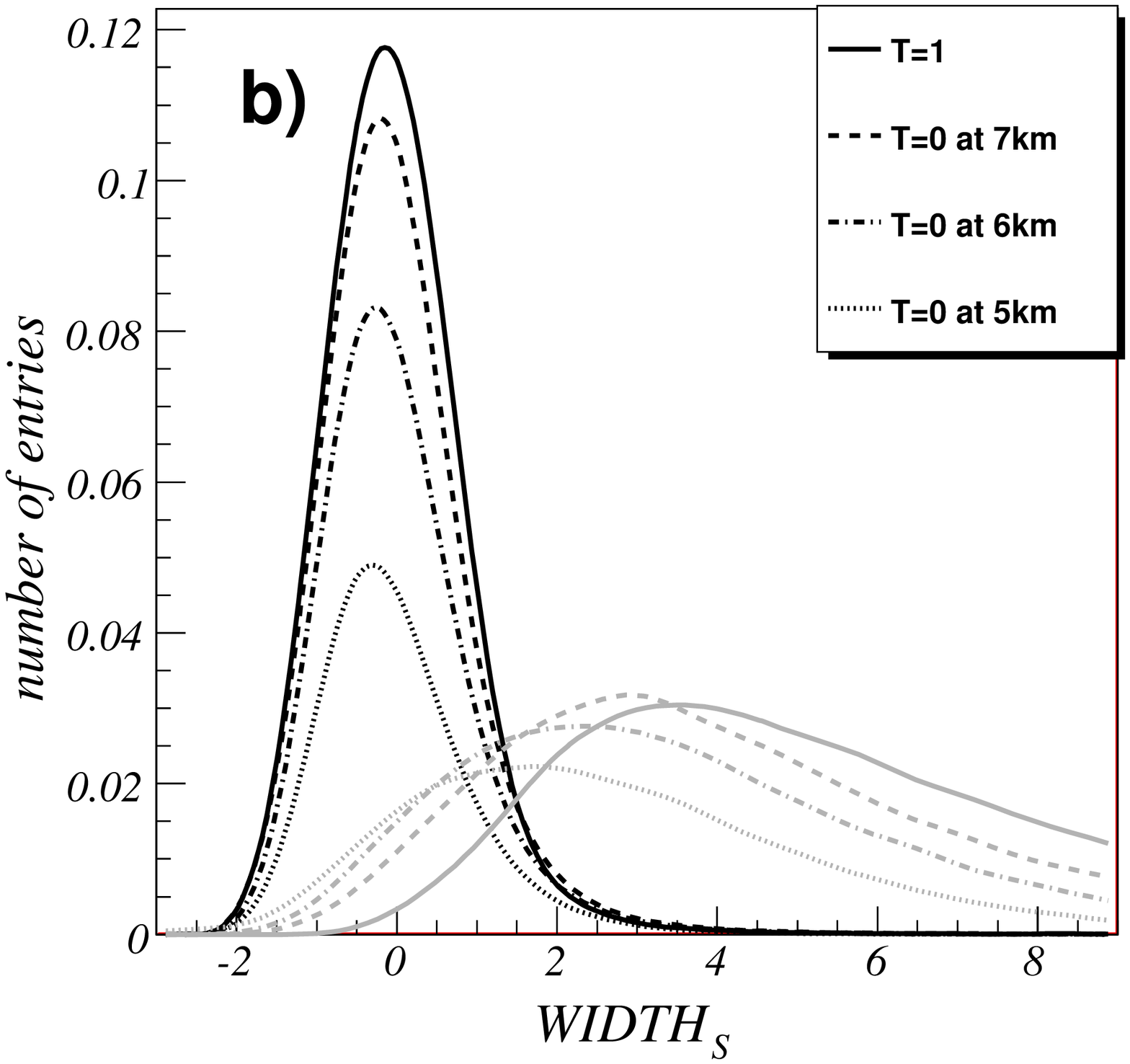}
\caption{The distributions of LENGTH$_{S}$ (a) and WIDTH$_{S}$ (b) for $\gamma$-ray initiated showers (black lines) and proton initiated showers (grey lines) in the case of fully opaque clouds at altitudes 7 km (dashed), 6 km (dashed-dotted) and 5km a.s.l. (dotted). The results for the clear sky simulations are shown by solid curves. Only images with the SIZE above 2000 photons are presented assuming the image cleaning level of 30 photons.}
\label{scalled}
\end{figure*}

We demonstrate the distributions of LENGTH$_{S}$ and WIDTH$_{S}$ for images of the $\gamma$-ray and proton initiated showers in the case of fully opaque clouds at different altitudes (see Fig.~5a,b). These distributions are made for images with SIZE above 2000 photons. It is easy to find out that the number of images with the SIZE above 2000 photons drop for the clouds at lower altitude.
These considered parameters still allow clear separation of the $\gamma$-ray induced showers in the presence of clouds. Note also that the distributions of WIDTH$_{S}$ allow better separation than distributions of LENGTH$_{S}$.
The $\gamma$/hadron separation applied in our work is based on the cuts in the WIDTH$_{S}$ and LENGTH$_{S}$. 
To demonstrate the selection efficiency we used the quality factor.  This factor is related to the significance of the measurement (the ratio of the $\gamma$-ray signal and the dispersion of the Poisson-like hadronic background) \cite{aha93}. The quality factor is defined as, 
$$Q_{\rm F}\equiv \frac{N^{\rm c}_{\gamma}/{N^{\rm all}_{\gamma}}}{\sqrt{N^{\rm c}_{\rm p}/N^{\rm all}_{\rm p}}},$$
where $N^{\rm all}_{\gamma}$,$N^{\rm all}_{\rm p}$ are the total numbers of events for chosen SIZE cut and $N^{\rm c}_{\gamma}$, $N^{\rm c}_{\rm p}$ are the numbers of events surviving the $\gamma$/hadron procedure. The lower limits on WIDTH$_{S}$ and LENGTH$_{S}$ were set to the value -2. The upper limits were optimized in the way to get the maximum value of $Q_{\rm F}$, for each investigated cleaning level, the altitude of the cloud, and its transmission. 
The quality factor versus the cloud transparency for the SIZE above 2000 photons and cleaning level of 45 photons
are shown in Fig.~6a. Black and grey lines correspond to different scaling parameters used to calculate WIDTH$_{S}$ and LENGTH$_{S}$. The scaling parameters, computed from the simulated $\gamma$-ray initiated images in the presence of clouds, were used to get $Q_{\rm F}$ (see black lines). The simulations of $\gamma$-ray initiated showers in case of clear sky conditions were used to calculate the scaling parameters. They are applied in the optimization of the quality factors (see grey lines). 
The $\gamma$-ray selection method is working much better in case of scaling made with the simulation of the real atmospheric conditions. 
For the cloud at 10 km a.s.l., the clear sky simulations can be used provided that the transparency of the cloud is larger then 0.4. 
However, for the cloud at lower altitudes, a suitable scaling should be used when analyzing data.
In this case, the quality factor increases with the transmission for the altitudes equal to 7, 6 and 5 km a.s.l.. In contrast, for 10 km, $Q_{\rm F}$ almost does not depend on the cloud transparency.

We also show the quality factors for images with SIZE above 10000 photons applying the image cleaning levels 30 photons and 60 photons (see black and grey curves in Fig.~6b). $Q_{\rm F}$ values are higher for the cleaning image on the level of 30 photons. It means that the presented lower cleaning level is sufficient to efficiently select $\gamma$-ray events. With the higher cleaning level, we observe the degradation of the quality factor, since less events survive the SIZE cut.

It has been shown in \cite{bern00} that a part of produced Cherenkov light, which is scattered by aerosols, is able to reach the ground. Clouds provide additional aerosols in the atmosphere. Therefore, higher level of NSB is required. This results in reduction of the $\gamma$/hadron separation efficiency. To confirm this, we performed additional simulations with different NSB. Figure 6c shows the quality factors for different NSB level. 
We assumed that, the NSB depends only on the cloud transmission and it linearly decreases with the transparency of the cloud. For the fully opaque cloud, we applied NSB larger by a factor 2.5 than in the case of clear sky simulations.
These calculations were obtained for the SIZE above 10000 and the cleaning level of 45 and 30 photons (see black and grey lines in Fig.~ 6c, respectively). We show that the higher NSB leads to significant degradation of the quality factors for the cloud transmission below 0.6, in the case the of cleaning on the level 30 photons. For higher transparencies, much smaller reduction of separation efficiency is expected. Using the higher cleaning level (45 photons), $Q_{\rm F}$ value significantly improves for all investigated transparencies and altitudes of clouds. We conclude that the effect of higher NSB level, due to the presence of clouds, can be partially compensated by using the higher level of image cleaning in the data analysis.

\begin{figure*}[t]
\vskip 4.5truecm
\includegraphics{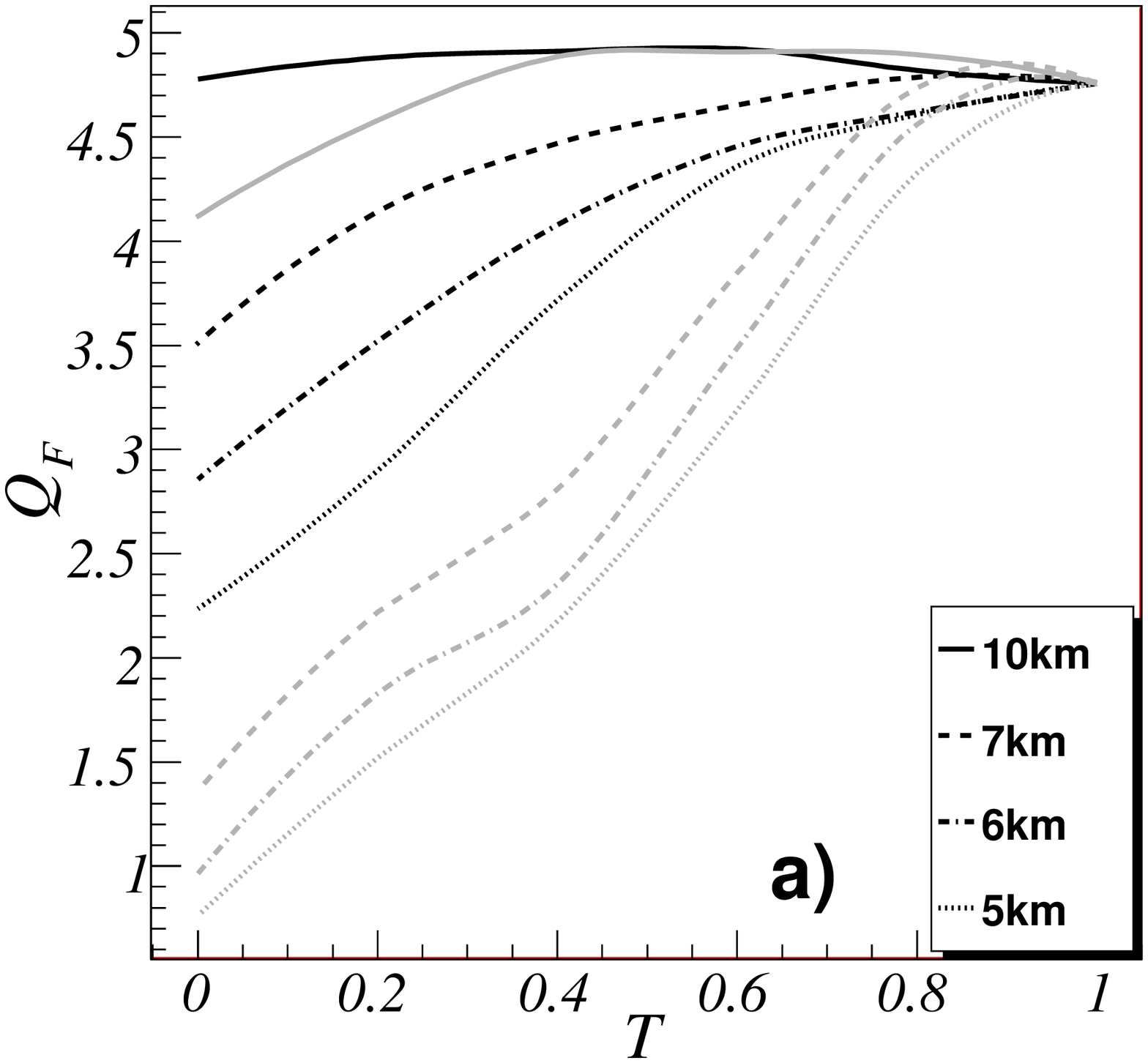}
\includegraphics{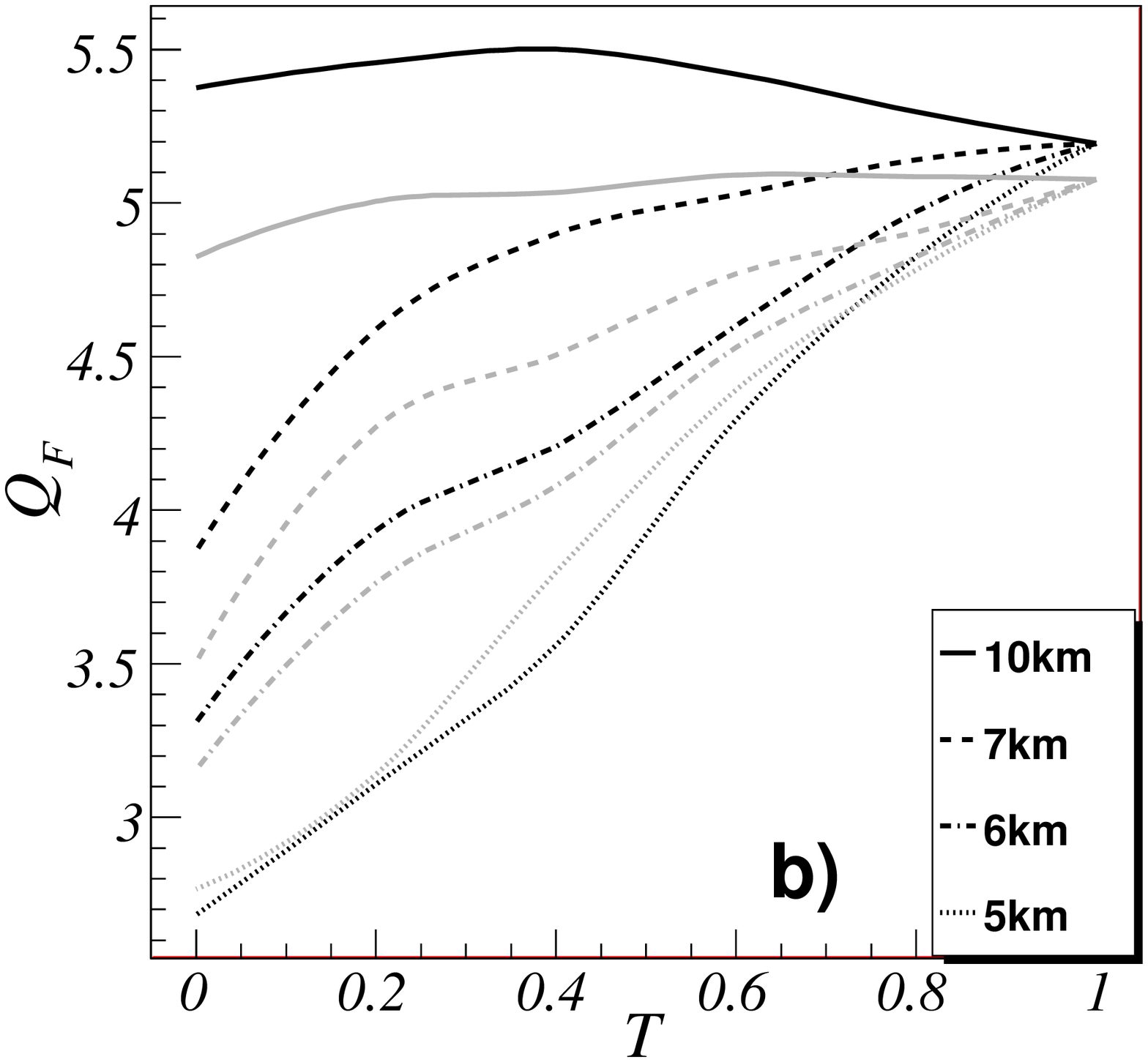}
\includegraphics{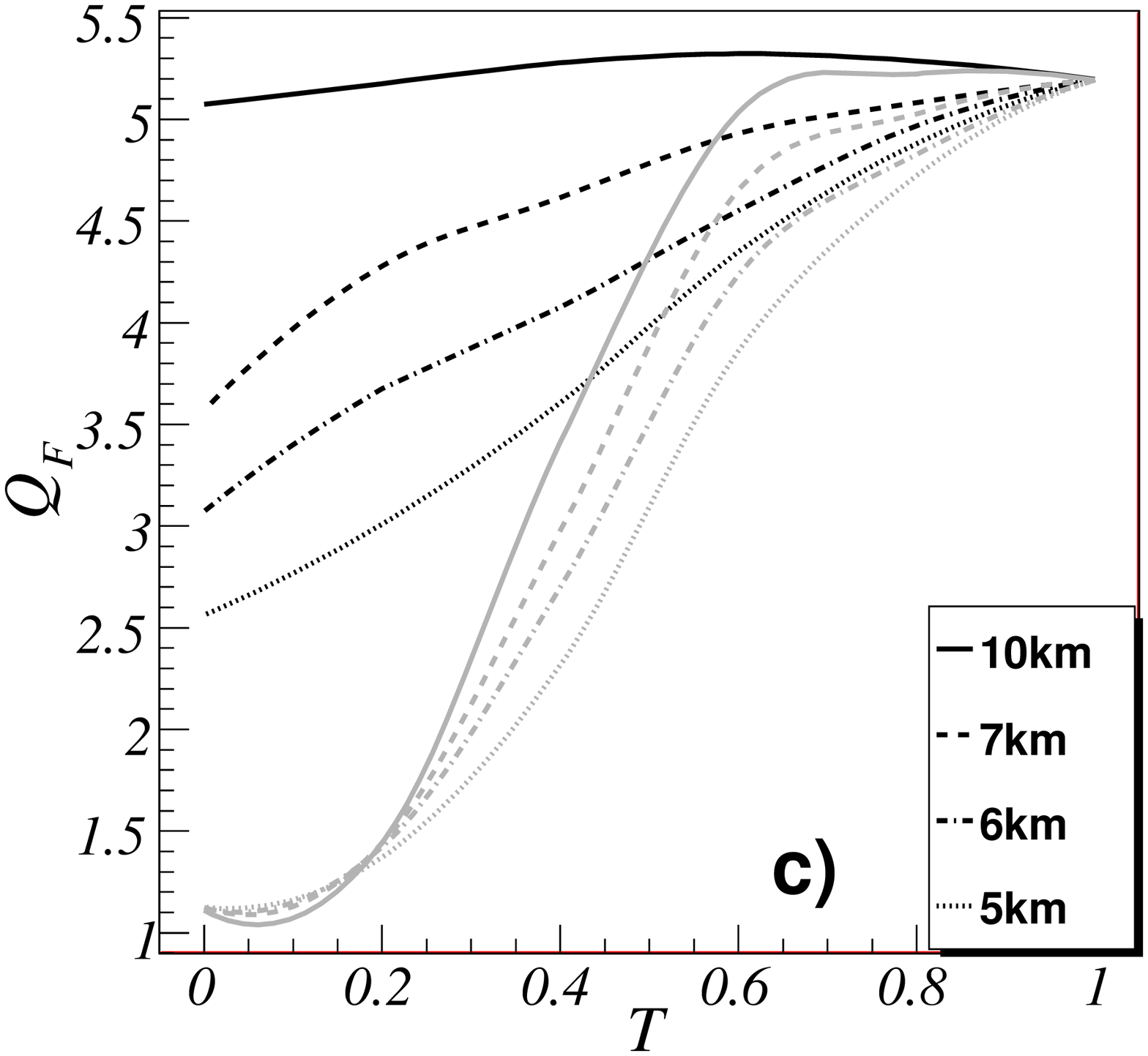}
\caption{The quality factor versus the cloud transmission for the cloud altitudes 10 km (solid lines), 7 km (dashed lines), 6 km (dashed-dotted lines) and 5 km a.s.l. (dotted lines).
{\bf a)} The comparison between results obtained using scaling factors from the simulation with the presence of the cloud (black lines) and from the clear sky simulations (grey lines) for the images with SIZE above 2000 photons. 
{\bf b)} The comparison between two cleaning levels 30 (black lines) and 60 photons (grey lines) for images with SIZE above 10000 photons. 
{\bf c)} The comparison between the results for the NSB dependent on the cloud transmission with the cleaning level of 45 photons (black lines) and the cleaning level of 30 photons (grey lines). The images with SIZE above 10000 photons are considered.}
\label{proton_par}
\end{figure*}

To check the impact of the limited camera FOV on the quality factors, we have considered only a part of the angular distributions of Cherenkov light on the ground, limited to 4$^{\circ}$ in x axis (corresponding to the full camera FOV of 8$^{\circ}$). We have repeated all calculations starting from the computation of the standard Hillas parameters. 
We show the comparison of the results obtained for the camera with limited FOV (grey lines) and results presented above (black lines), see Fig.~7a,b. Both plots are obtained with the cleaning level of 30 and the SIZE above 10000 photons. Surprisingly, the quality factors for camera with limited FOV (Fig.~7a) are slightly better for almost all investigated transmissions and altitudes of the cloud. To explain this fact, we demonstrate the fraction of the $\gamma$-ray images, which survives the selection procedure as a function of the cloud transparency (Fig. 7b). It is clear, that more $\gamma$-ray events are recognized for the camera with unlimited FOV (black lines) than for the camera with limited FOV (grey lines). We have checked, that the same tendency is observed for the fraction of protons which survive cuts. The differences between cameras with unlimited and limited FOV are greater for the proton initiated showers than for the $\gamma$-ray initiated showers. As a consequence, we obtain larger quality factor for the camera with limited FOV. It proves that the $\gamma$/hadron separation, based on the optimization of the WIDTH$_{S}$ and LENGTH$_{S}$ cuts, is possible at the very high energies in the presence of clouds even in case of the camera with FOV limited to 8$^{\circ}$. 
The smaller fraction of images survive selection criteria for the camera with fixed size because for some events a part of the image is not recorded. This results in a smaller SIZE value. Also the Hillas parameters can change.
Both effects are important in optimization of the $\gamma$/hadron separation. On one hand, a part of events do not survive SIZE cut. On the other hand, the leakaged images of $\gamma$-ray initiated showers are more similar to images of the proton showers.
The impact of the camera edge, on the fraction of identified $\gamma$-rays with SIZE above 10000 photons is demonstrated in Fig.~7b. The fraction of identified $\gamma$-ray events for the camera with limited FOV is approximately 8$\%$ lower than for the camera with unlimited FOV in the case of the clear sky simulations.

\begin{figure*}[t]
\vskip 6.truecm
\includegraphics{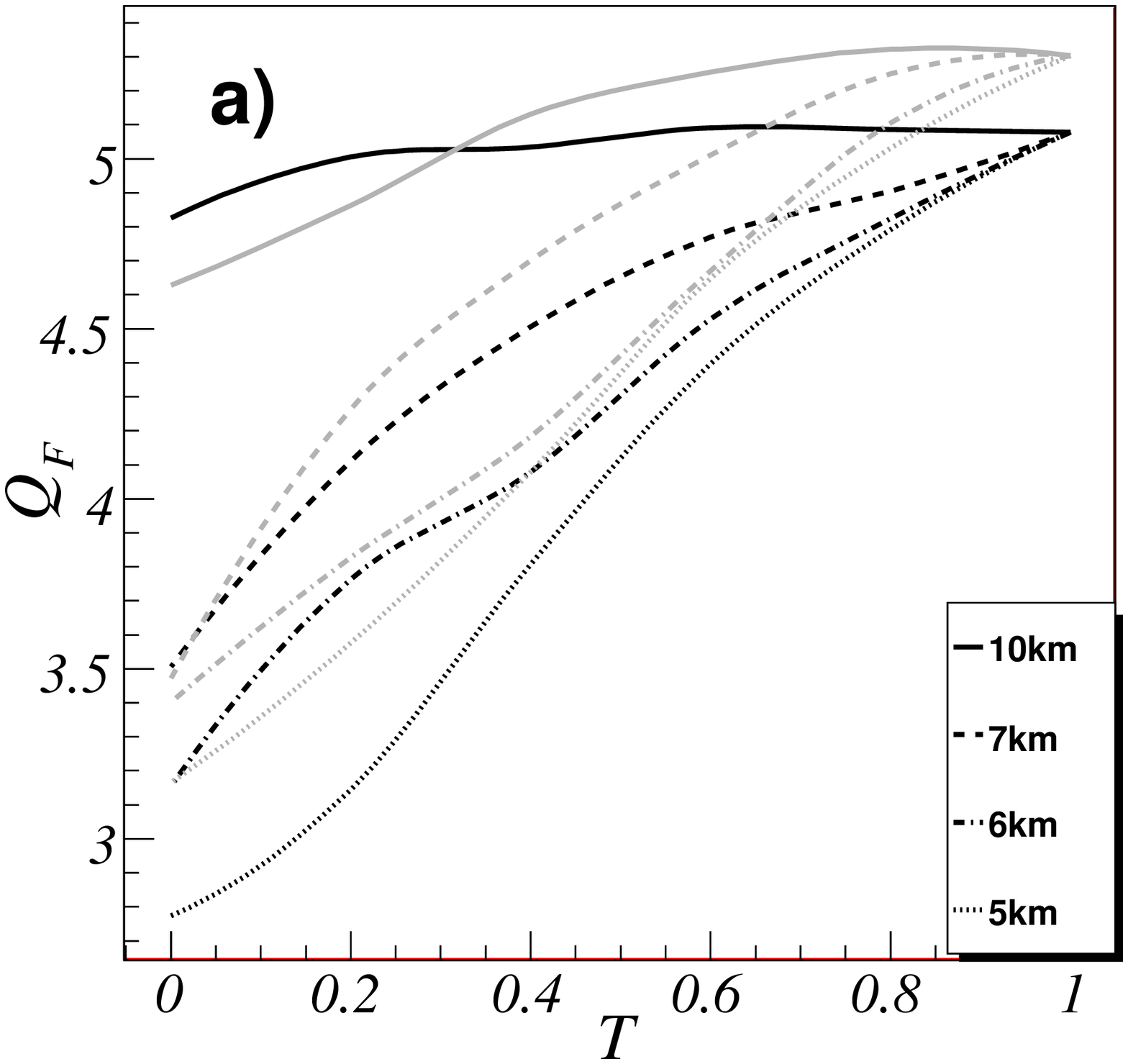}
\includegraphics{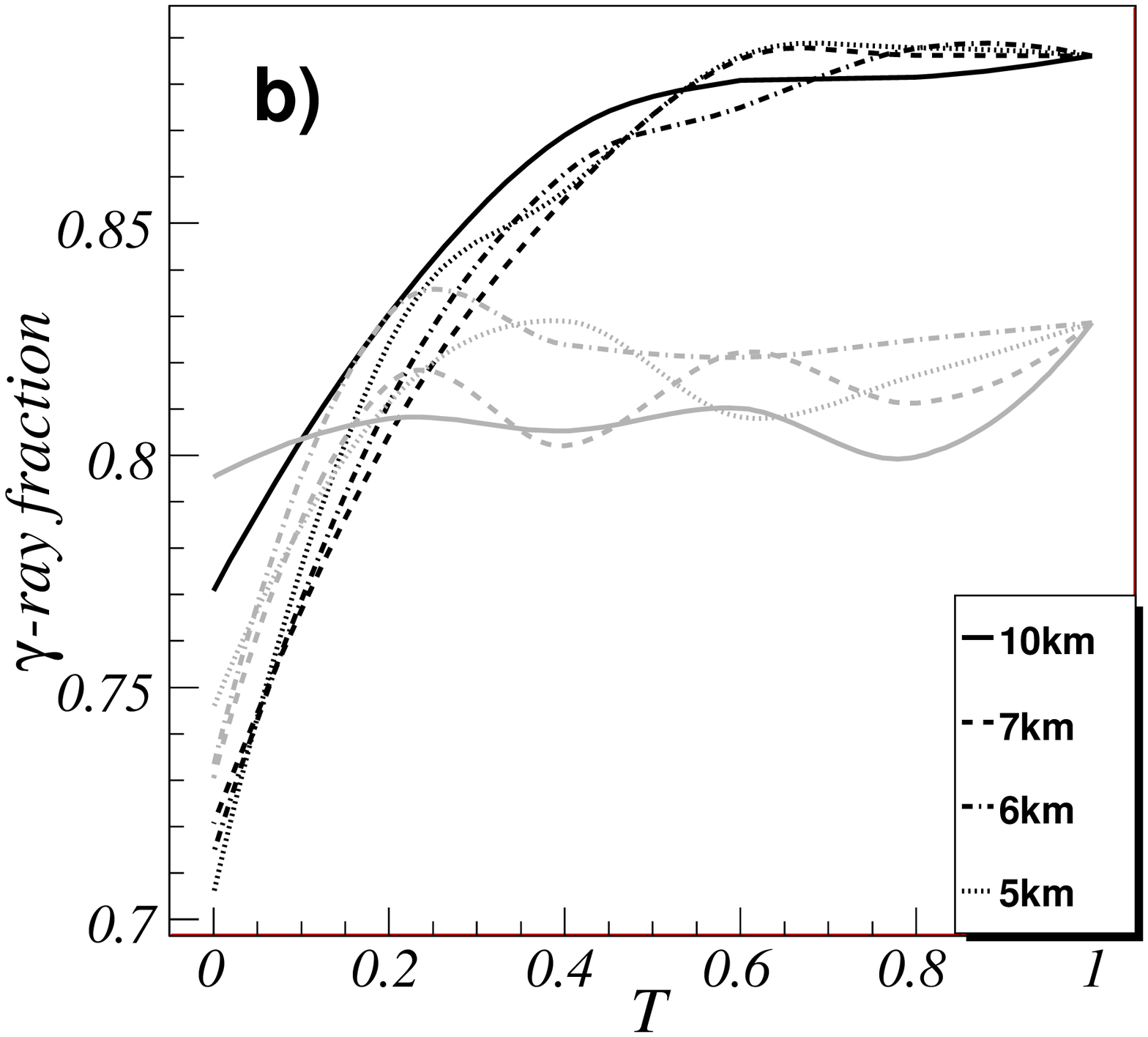}
\caption{The quality factor ({\bf a}) and the fraction of identified $\gamma$-ray initiated showers ({\bf b}) versus the cloud transmission parameter for the unlimited camera FOV (black lines) and for the camera FOV limited to 8$^{\circ}$ (grey lines). The clouds at the altitudes 10 km (solid lines), 7 km (dashed lines), 6 km (dashed-dotted lines) and 5 km a.s.l.(dotted lines) are analysed. The results are obtained for the SIZE above 10000 and the cleaning level of 30 photons.}
\label{qul_4deg}
\end{figure*}

\subsection{Possible implications for stereoscopical observations and the energy reconstruction} 

 There are two types of the observations with the Cherenkov telescopes. In the first (so called mono mode), the single IACT is detecting one image of the EAS. In the second (so called stereo mode), a system of telescopes records at least two images of the same shower. The stereo mode has advantages in respect to the mono mode of the observations, since showers are better reconstructed \cite{hint2009}. 

Using at least two recorded images allows to replace WIDTH$_{S}$ and LENGTH$_{S}$ parameters by mean scaled WIDTH and mean scaled LENGTH \cite{Kon99}. As a result, more efficient hadron rejection is obtained in the stereo mode. All images of the same shower can be changed due to the cloudiness. The efficiency of the $\gamma$/hadron separation, based on the WIDTH$_{S}$ and the LENGTH$_{S}$, declines in the presence of clouds. Therefore, we expect similar worsening of the $\gamma$/hadron separation also for the observations with the system of Cherenkov telescopes due to the cloudiness. 

The reconstructed shower direction depends on the direction of the major axis of the image for observations with a single telescope. In the stereo mode the intersection of the directions, defined by the major axes of the images recorded in each camera, determines the shower direction. Therefore, the direction of the shower is more accurate with the stereo mode of the observation. The presence of clouds do not influence determination of the major axis (or axes) of the image recorded by camera with unlimited FOV for very high energies. Therefore, the capability of the $\gamma$-ray selection, based on the parameters describing the orientation of the image, should be similar for the case of clear sky observation and the observation with the cloudiness.

In the case of mono observations at very high energies, the reconstructed energy increases with both the SIZE parameter and DIST parameter. 
In the presence of clouds the image SIZE decreases but the DIST parameter can increase. First effect causes the bias in the reconstructed energy towards lower values. This bias can be only partially compensated by the second effect. We have shown that the shift in the DIST distributions of the $\gamma$-ray showers is significant for clouds with low transmission only.  

The position of the shower core is better determined in the stereo observations. For this reason, the primary energy of the $\gamma$-ray is more accurately estimated than for the single IACT. In the presence of clouds, the shower reconstruction becomes worse also with the stereo mode of the observations. For this reason, the the reconstructed energy is biased towards lower values \cite{nolan10}.
Reconstruction of the primary energy of the $\gamma$-ray is based on the MC results for a particular single telescope or system of IACTs. 
The MC simulations have to take into account the presence of clouds in order to avoid a bias in the reconstructed energy.

\section {Conclusions}

We investigate the impact of the presence of clouds on the Cherenkov photon densities and the parameters of the images detected by IACT for the $\gamma$-ray and proton initiated showers in the atmosphere assuming that the camera of the telescope has unlimited or limited to 8$^{\circ}$ field of view.
We have shown that the impact of fully opaque clouds on the expected Cherenkov photon density on the ground is smaller than the influence of the limited to 8$^{\circ}$ camera FOV for the shower impact parameters larger than the hump position.
The presence of clouds affects both $\gamma$-ray and proton images seen by Cherenkov telescopes. At very high energies i.e. above a few TeV, the changes of images are negligible for cloud altitudes above 10 km a.s.l., since those altitudes are above the shower maximum.
The images registered in the presence of the clouds at lower altitudes are shifted in the camera plane. They are less concentrated in comparison to the events seen with the clear sky conditions. 
With the presence of clouds, the DIST distribution moves to the larger values. This effect decreases with the cloud transmission for the fixed cloud altitude. For clouds with the transparencies larger then 0, the maximum in the LENGTH distribution shifts to larger values with dropping transmission of the cloud. 
The impact of clouds on the WIDTH distribution is similar for $\gamma$-ray and proton initiated showers. The WIDTH distributions become wider when the cloud transmission decreases. Due to geometrical effect, lower values of the LENGTH and larger values of the WIDTH parameters are expected for the fully opaque clouds in comparison to the results obtained with the clear sky conditions.

The $\gamma$/hadron separation, based only on the cut optimization in scaled WIDTH and LENGTH parameters, is possible. However, three important effects have to be taken into account. At first, the scaling factors should be calculated from the adequate simulations, which include atmospheric conditions. If the scaling factors are calculated from the simulations which neglect absorption of Cherenkov photons in the cloud, then the efficiency of the $ \gamma$-ray selection procedure is significantly reduced. In the worst case, it is reduced by a factor of $\sim$4 in comparison to the proper scaling results obtained for the fully opaque clouds at the altitude of 5 km a.s.l.. 
At second, if the level of the image cleaning is too high, then the efficiency of the $\gamma$/hadron separation is slightly worse in comparison to proper cleaning. 
At third, the higher NSB is expected in the presence of clouds. This leads to the degradation of the $\gamma$/hadron separation efficiency. This last effect can be partially compensated by using higher level of image cleaning.\\
We have shown, that the quality factors could be slightly better for the telescope with 8$^{\circ}$ camera FOV than for the camera with
unlimited FOV. The camera edge effects reduce the fraction of identified $\gamma$-ray events by smaller factor than the fraction of proton images surviving cuts.

We conclude that the large mirror Cherenkov telescopes, equipped with the camera with large FOV, should allow observation of $\gamma$-ray showers at very high energies with the presence of high altitude clouds. Therefore, the effective observation time of IACTs could be significantly increased.

\ack

This work was supported by Polish Grant from Narodowe Centrum Nauki 2011/01/M/ST9/01891

\end{document}